\documentclass[aps,pre,preprint,groupedaddress,showpacs]{revtex4}
\usepackage{graphicx}
\usepackage{amsmath}
\begin{document}
\title{Persistent entanglement due to helicity conservation 
in excitable media}
\author{Jos\'e L. Trueba and Manuel Array\'as}
\affiliation{\'Area de Electromagnetismo, Universidad Rey Juan
Carlos, Camino del Molino s/n, 28943 Fuenlabrada, Madrid, Spain}

\begin{abstract}
This work addresses the topic of knotted stable structures in 
excitable media. These structures appear in a wide variety of situations, 
such as cardiac fibrillation, chemical reactions, etc. Entangled 
curves have been found in numerical computations of the equations 
that describe excitable media. They present an unusual stability. 
An explanation for this behaviour has been an open question.
In the present work we introduce for the first time the meaning 
of the helicity in an excitable media as a new tool to study 
the stability of these systems. The helicity is related to 
the total entanglement of the system. We have studied how 
the helicity is conserved or lost through the walls of the medium 
and shown that these behaviours are dominated by the boundary conditions, 
so the distortion of these conditions could lead to the disappearance 
of the structures.
\end{abstract}

\date{\today}
\maketitle

There has been experimental reports on stable
structures (rotors) in cardiac muscle that appear in cardiac
fibrillation \cite{r1}, which may support the ideas of Arthur T.
Winfree \cite{r2} on the relation between electrical wave
propagation in the heart and some stable structures appeared in
excitable media. One of the most interesting aspects of excitable
media is that they support vortices, called spiral or scroll
waves. Spiral waves are characterized by the fact that they rotate
around a topological point defect called phase singularity or
rotor. Since spiral waves are the cross sections of a scroll wave,
the phase singularity can be interpreted as the cross section of a
filament in three dimensions. These curves can be linked or
constitute knots. The important fact of the phase singularities is
that they are the focus or organizing centres of the excitation of
the complete medium, and that their stability has been numerically
proof for some particular cases \cite{r3}. The stability of phase
singularities has been studied locally \cite{r4}. However,
definitive conclusions on the behaviour of the singular filaments
have not been obtained. An alternative approach to this problem
could come from the study of non-local mechanisms that allow
certain topological invariants of the organizing centres to be
conserved. Preliminary studies on topologically non-trivial field
configurations have appeared \cite{r5,r6}. Examples are the work
by Berry and Dennis \cite{r7} on phase singularities in the
Helmholtz equation or studies on the stability of ball lightning
\cite{r8,r9}. Here we use a new approach to the stability of phase
singularities in excitable media. We define the helicity of the
excitable medium and we study its meaning in relation to the total
entanglement of the system. We consider the time behaviour of the
FitzHugh-Nagumo model, showing that the persistence of
entanglement depends strongly on the conservation of the helicity.
This observation could be of some utility to develop new methods
of controlling the appearance of rotors by acting on the system
boundaries in experimental situations.

Excitable media describe in good approximation some properties of
certain chemical reactions \cite{r10}, cardiac arrhythmias
\cite{r11}, etc. The simplest mathematical models for propagation
in three-dimensional excitable media include two state variables
$u$ and $v$ that satisfy the equations
\begin{eqnarray}
\frac{\partial u}{\partial t} = \frac{1}{\varepsilon} \, f (u,v) +
D_{u} \nabla^2 u \, , \nonumber \\
\frac{\partial v}{\partial t} = \varepsilon \, g (u,v) + D_{v}
\nabla^2 v \, . \label{eq1}
\end{eqnarray}
Here $u$ is the excitation variable, $v$ is the inhibition
variable and $\varepsilon$ is a small parameter. At one particular
point in the spatial domain ${\cal D}$ in which the variables $u$
and $v$ are defined, both quantities evolve in time according to
(in general, nonlinear) functions $f(u,v)$ and $g(u,v)$
respectively. The coupling between close points in the medium
occurs due to diffusion terms with coefficients $D_{u}$ and
$D_{v}$. When the system parameters have specific range of values,
the excitation of a point of the system propagates as a shockwave,
with a propagation velocity given by the value of $\varepsilon$
and the diffusion coefficients \cite{r12}.

In order to study the topology of singular filaments in excitable
media, we define a vector field in which the field lines coincide
with the intersections of level surfaces of $u$ with level
surfaces of $v$, since these intersections include the phase
singularities that organize the complete medium around them. The
linking number is a measure of the extent to which the field lines
of a divergence-free vector field curl themselves around one
another, i.e. of the helicity of the field as was defined by
Moffatt \cite{r13} in 1969.

Consider the situation of an excitable media, given by a pair of
real scalar fields $(u,v)$ defined in a three-dimensional spatial
domain ${\cal D}$, that satisfy the system of equations
(\ref{eq1}). Our aim is to describe the unusual stability of
certain configurations from a new global point of view, taking
into account the linkage of the curves obtained from the
intersections of level surfaces of $u$ with level surfaces of $v$.
Suppose that we are interested in a physical situation in which
$u$ takes values between $u_{min}$ and $u_{max}$, and $v$ takes
values between $v_{min}$ and $v_{max}$. We define new variables
$U$ and $V$ from $u$ and $v$ through linear scaling, in such a way
that $U$ and $V$ satisfy $0 \leq U^2 + V^2 \leq 1$. Note that the
level surfaces of $u$ and $v$ coincide with the level surfaces of
$U$ and $V$ respectively, since the change is linear. Now we
define the variables $p= \sqrt{U^2 + V^2}$, $q=ArcTan{(V/U)}$. A
complex scalar field   that describes the excitable medium is then
given by
\begin{equation}
\phi = \sqrt{\frac{1-p}{p}} \, e^{iq} \, . \label{eq3}
\end{equation}
The level curves of $\phi$ are the intersections of the surfaces
of constant $u$ and the surfaces of constant $v$ at any time. We
now define a vector field given by
\begin{equation}
{\boldsymbol{\Omega}} = \frac{\nabla \phi \times \nabla
\bar{\phi}}{2 \pi i \left( 1 + \bar{\phi} \phi \right)^2} \, .
\label{eq4}
\end{equation}
The field lines of ${\boldsymbol{\Omega}}$ coincide with the level
curves of $\phi$ by definition. In equation (\ref{eq4}), $i$ is
the imaginary unit and $\bar{\phi}$ is the complex conjugate of
$\phi$. The particular definition (\ref{eq4}) has an interesting
bonus. If the scalar is a map $\phi : S^3 \rightarrow S^2$, these
maps yield knots and can be classified in homotopy classes
characterized by the integer value of the Hopf index $H(\phi )$,
which gives a measure of the linkage or entanglement of the level
curves of the map. Since the vector field ${\boldsymbol{\Omega}}$
defined by equation (\ref{eq4}) is divergence-free, a vector
potential ${\boldsymbol{\Psi}}$ can be found so that
${\boldsymbol{\Omega}} = \nabla \times {\boldsymbol{\Psi}}$. The
Hopf index $H (\phi )$, that is a topological invariant of the map
$\phi : S^3 \rightarrow S^2$, can be then written as the integral
\begin{equation}
H (\phi) = \int \left( {\boldsymbol{\Psi}} \cdot
{\boldsymbol{\Omega}} \right) d^3 r \, . \label{eq6}
\end{equation}
This integral is known in fluid and plasma physics to be the
helicity of the vector field ${\boldsymbol{\Omega}}$, a global
measure of the linkage of the force lines of
${\boldsymbol{\Omega}}$. Consequently, if we define the helicity
of the configuration, at any time, as in equation (\ref{eq6}),
this quantity will inform us about the global linkage of the
intersections of the level surfaces of $u$ and $v$ at any time
and, more important, if this quantity is conserved during the
evolution of the system given by equations (\ref{eq1}), then the
{\emph global} topology of the initial configuration is preserved,
constituting a strong source of stability of the system. In
general, $\phi$ will not correspond to a real map from $S^3$ to
$S^2$, so its helicity will not be equal to a Hopf index. This may
happen if, for example, the domain ${\cal D}$ is not infinite but
a box with finite edges. Then the value of the helicity will be a
real number instead an integer one, but its meaning is always
related to the global linkage of the intersections of the level
surfaces of $u$ and $v$.

An interesting observation can be noted here on the vector field
defined from the $(u, v)$ configuration through equations
(\ref{eq3}, \ref{eq4}): it is parallel to the vector field $\nabla
u \times \nabla v$, whose maximum value is used by many authors to
detect the vortex that organizes the medium \cite{r6}. In terms of
the global scheme described in this work, the explanation of this
fact is related with high values of the helicity density in
regions where the density of linked lines is also high.

Suppose that a particular initial configuration has been given, in
which the initial helicity has a non-zero value. When the system
evolves in time according to equations (\ref{eq1}), due to the
presence of diffusion there will be reconnections of the lines
given by intersections of level surfaces, and the value of the
helicity will change in general. However, as we will see, there
are situations in which the helicity remains constant, or it
changes very slowly with time compared to the characteristic time
of the system. In these situations, the non-zero value of the
helicity reflexes an unusual stability of the configuration that
can explain important numerical, or even experimental,
observations. The time variation of the helicity, from equation
(\ref{eq6}), is
\begin{equation}
\frac{d H (\phi)}{dt} = \int_{S} {\boldsymbol{\Psi}} \cdot \left(
{\bf u}_{N} \times \frac{\partial {\boldsymbol{\Psi}}}{\partial t}
\right) d S \, . \label{eq7}
\end{equation}
Here, $S$ is the boundary of the three-dimensional domain ${\cal
D}$ and ${\bf u}_{N}$ is a unit vector orthogonal to the surface
$S$ at each point. The integral (\ref{eq7}) has to be computed on
the boundary of the domain. In equation (\ref{eq7}) we have
\begin{equation}
\frac{\partial {\boldsymbol{\Psi}}}{\partial t} = \frac{-1}{2 \pi
i \left( 1 + \bar{\phi} \phi \right)^2} \left( \frac{\partial
\bar{\phi}}{\partial t} \nabla \phi - \frac{\partial
\phi}{\partial t} \nabla \bar{\phi} \right) \, . \label{eq8}
\end{equation}
Expression (\ref{eq7}) can be used to investigate in what cases
the helicity is constant during the evolution of the system and
how it changes in other cases. A common feature is that the
conservation of the helicity depends almost only on the boundary
conditions, so that the stability of the system can be perturbed
numerically by acting only on the surfaces of the medium. This
observation could be of some utility in experimental situations.

Now let us examine some typical boundary conditions. First, if the
domain is the complete $R^3$ space and the fields are taken
initially so that they are zero at infinity, then the vector field
$\partial {\boldsymbol{\Psi}}/\partial t$ will always be zero at
the boundaries and the helicity will be conserved in time
according to equation (\ref{eq7}). This is the case of the
electromagnetic knots in vacuum \cite{r14,r15,r16,r17,r18,r19}. A
similar situation happens if Dirichlet boundary conditions are
imposed in all the boundaries of a finite box provided the scalars
$u$ and $v$ are smooth functions of space and the time evolution
is also smooth. However, in numerical computations of excitable
media, Neumann (null flux) and periodic boundary conditions are
mostly used. As an example, consider that the domain ${\cal D}$ is
a grid in $R^3$ in which the spatial coordinates $(x, y, z)$ are
confined to the range $-L \leq x, y, z \leq L$, $L$ being a
certain length, and suppose that Neumann boundary conditions are
applied to the $x$ and $y$ directions, and periodic boundary
conditions are applied to the $z$ direction. In this example,
there will not be loss of helicity through the $z$ direction, but
helicity will be lost through the $x$ and $y$ directions in an
amount that depends on the parameters of the model that one is
computing. Writing $\partial {\boldsymbol{\Psi}}/\partial t$ in
terms of $u$ and $v$ through equations (\ref{eq3}, \ref{eq8}), the
term in equation (\ref{eq7}) is proportional to ${\bf u}_{N}
\times (\partial_{t} v \nabla u - \partial_{t} u \nabla v )$.
Taking into account the contributions of both the faces $z=- L$
and $z=L$ in which the field is periodic, this is equal to zero,
so that the helicity is conserved in the directions in which
periodic boundary conditions are applied. In the faces in which
Neumann boundary conditions are applied, the term ${\bf u}_{N}
\times (\partial_{t} v \nabla u - \partial_{t} u \nabla v )$ is
not zero but it depends on the parameters of the model because
$\partial_{t} u$ and $\partial_{t} v$ depend on them through the
evolution equations (\ref{eq1}).

We will give an example of the helicity conservation in the
FitzHugh-Nagumo model. This model set a paradigma which allows a
geometrical explanation of important biological phenomena related
to neuronal excitability and spike-generating mechanism. It have
been extensively studied \cite{r6} and it was conjectured that
persistent solutions called organizing centres might exist in
three dimensions for excitable media in which two-dimensional
vortices are embedded into three dimensional space forming knotted
vortex rings. In order to prove the existence of these solutions,
a theoretical framework based on local analysis involving
effective models of short-range repulsive forces between vortex
cores was proposed, but only certain limiting cases of slight
curvature and twist of vortex lines had partial results
\cite{r4,r12,r20}. Finally to address the fundamental issue of the
existence of stable knots, numerical investigations were
performed. Here we will analyse again the FitzHugh-Nagumo model
but at the light of the global stability provided by helicity
conservation. The FitzHugh-Nagumo equations are given by
\begin{eqnarray}
\frac{\partial u}{\partial t} = \frac{1}{\varepsilon} \, \left( u
- u^3 /3 - v \right) + \nabla^2 u \, , \nonumber \\
\frac{\partial v}{\partial t} = \varepsilon \, \left( u + \beta -
\gamma v \right) \, . \label{eq9}
\end{eqnarray}
Here $u$ represents the electric potential and $v$ the recovery
variable associated with membrane channel conductivity. We choose
the constants appearing in equation (\ref{eq9}) to have the values
\cite{r6} $\varepsilon = 0.3$, $\beta = 0.7$ and $\gamma = 0.5$.
We discretized the equations (\ref{eq9}) with finite differences
on a cubic domain of size $2L$ with a uniform cubic grid of
spacing $h$. For the Laplacian operator we use a second order
accurate finite difference approximation which is symmetrical up
to third order. We have to keep in mind the stability criteria
about $t < h^2/2D$ when setting the spatial mesh.

In order to test the conservation of helicity we enforce $u =
-1.03279$, $v = -0.66558$ at the boundaries, which are the
equilibrium values of the system. We take initial conditions with
nonzero helicity,
\begin{eqnarray}
u (x,y,z,0) = \lambda_{1} \frac{2xz + y ( x^2 + 
y^2 + z^2 -1 )}{(x^2 + y^2 + z^2 +1 )^2} - 0.4  \, , \nonumber \\
v (x,y,z,0) = \lambda_{2} \frac{2yz - x ( x^2 + y^2 + z^2 -1
)}{(x^2 + y^2 + z^2 +1 )^2} - 0.4 \, . \label{eq10}
\end{eqnarray}
Here, $\lambda_{1} = \sqrt{2}$ and $\lambda_{2} = 1/\sqrt{2}$
respectively, which are used to cover the range of the
excitation-recovery loop in $(u, v)$ space \cite{r6} for the
ordinary differential equation part of equations (\ref{eq9}). Now
for testing the conservation of helicity, we will solve the system
and plot the intersection of level surfaces. Each level surface of
constant $u$ can intersect another surface of constant $v$ in a
curve, few curves or an empty set. If helicity is conserved and
the initial conditions are chosen in such a way that intersection
of level surfaces are knotted with entanglement or linking number
different from zero, we can find as time evolves that there must
be intersection curves linked all the time.

\begin{figure}
\centering
\includegraphics[width=0.6\textwidth]{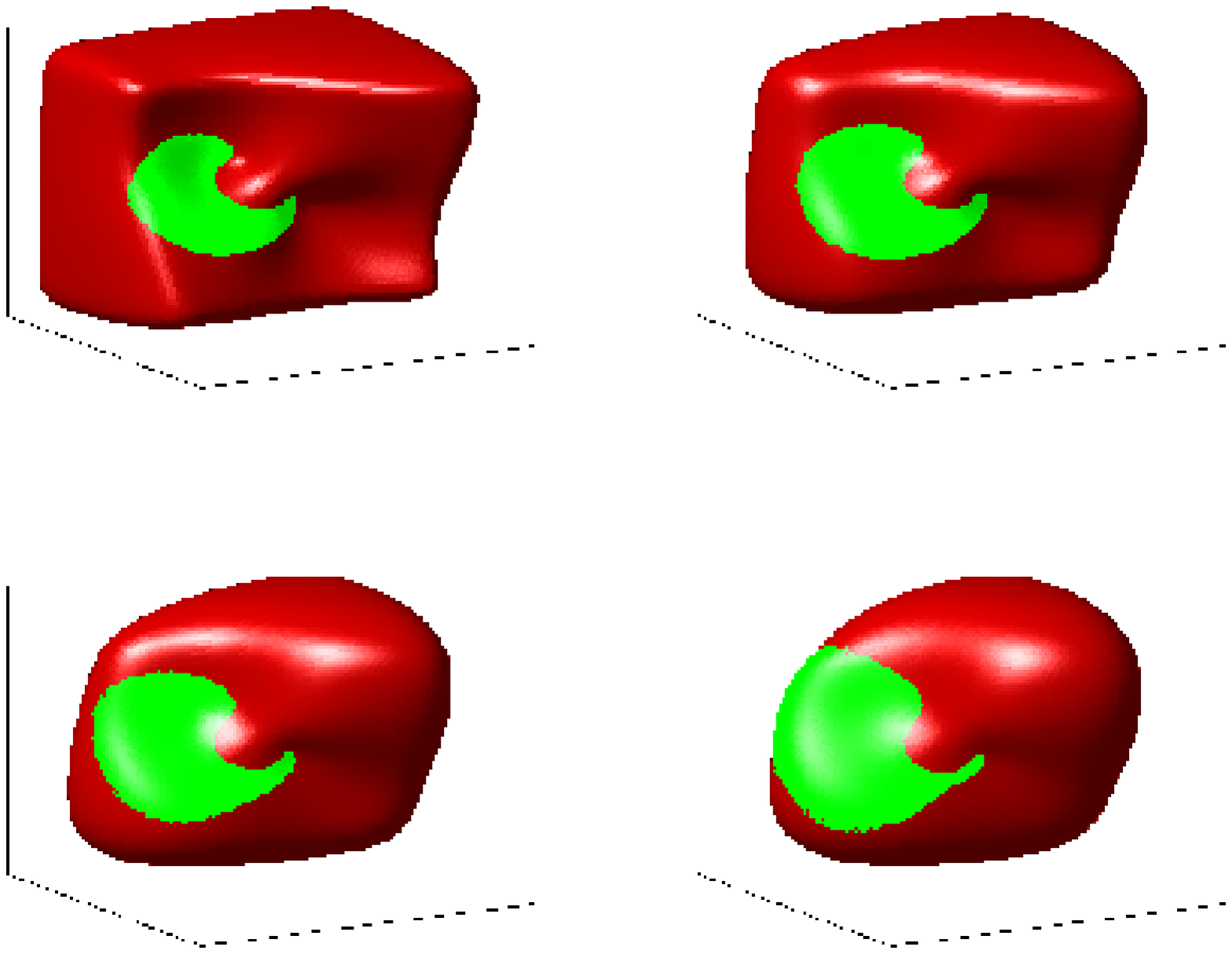}
\caption{{\bf Evolution of surface levels}. The figure shows the
evolution of two level surfaces, u = -0.7 (red) and v = -0.1
(green). From left to right and top to bottom, the snapshots
correspond to 0.2, 0.4, 0.6 and 0.8 instants of simulation time.}
\label{fig1}
\end{figure}

\begin{figure}
\centering
\includegraphics[width=0.3\textwidth]{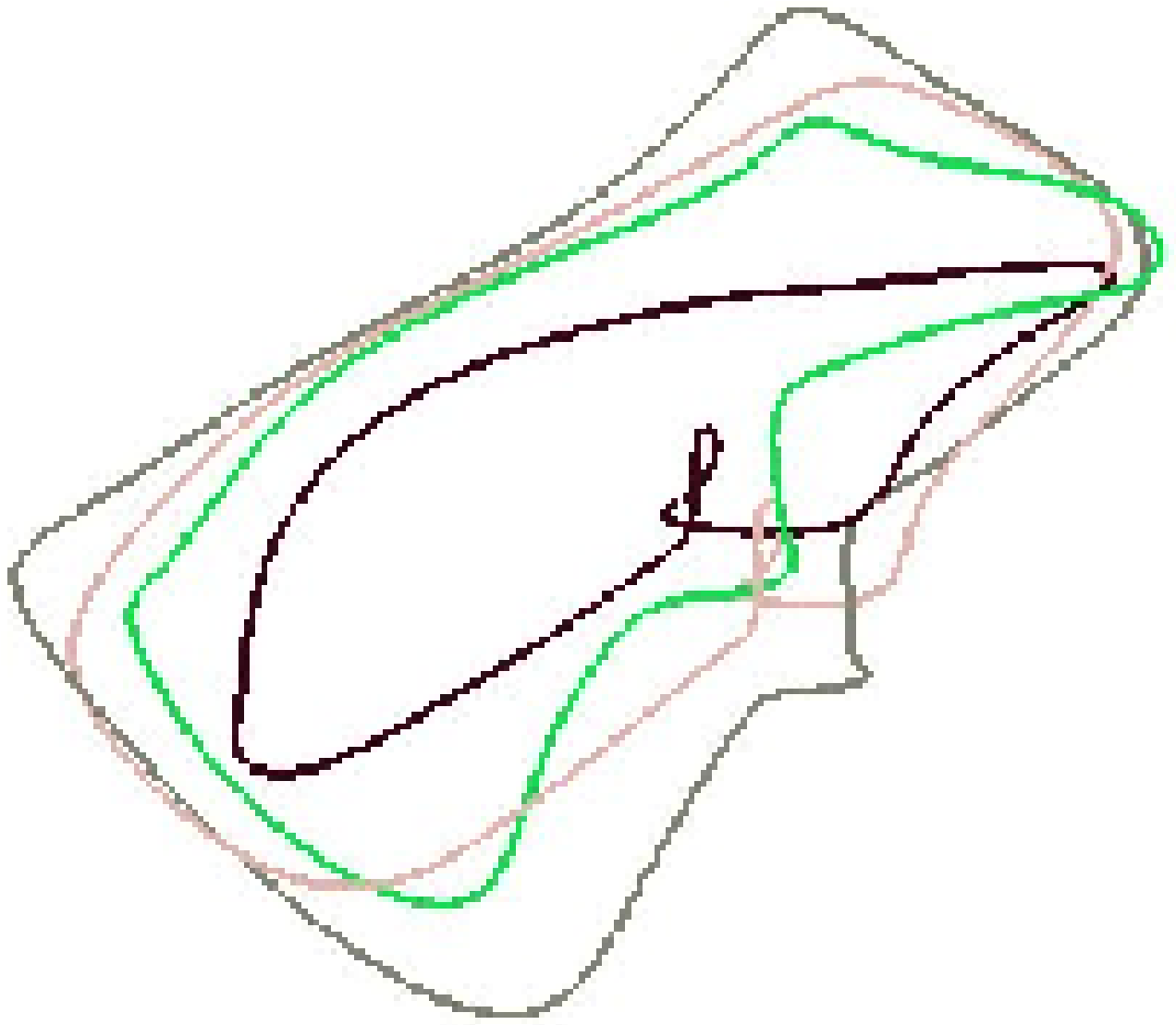}
\includegraphics[width=0.3\textwidth]{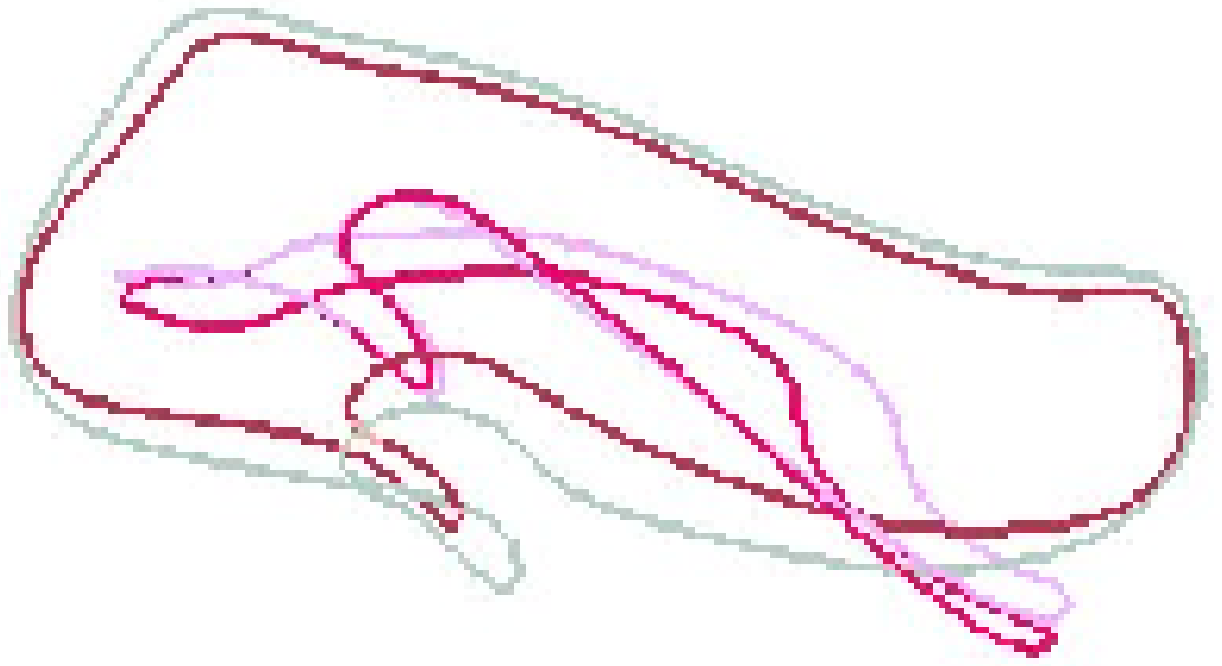}
\includegraphics[width=0.3\textwidth]{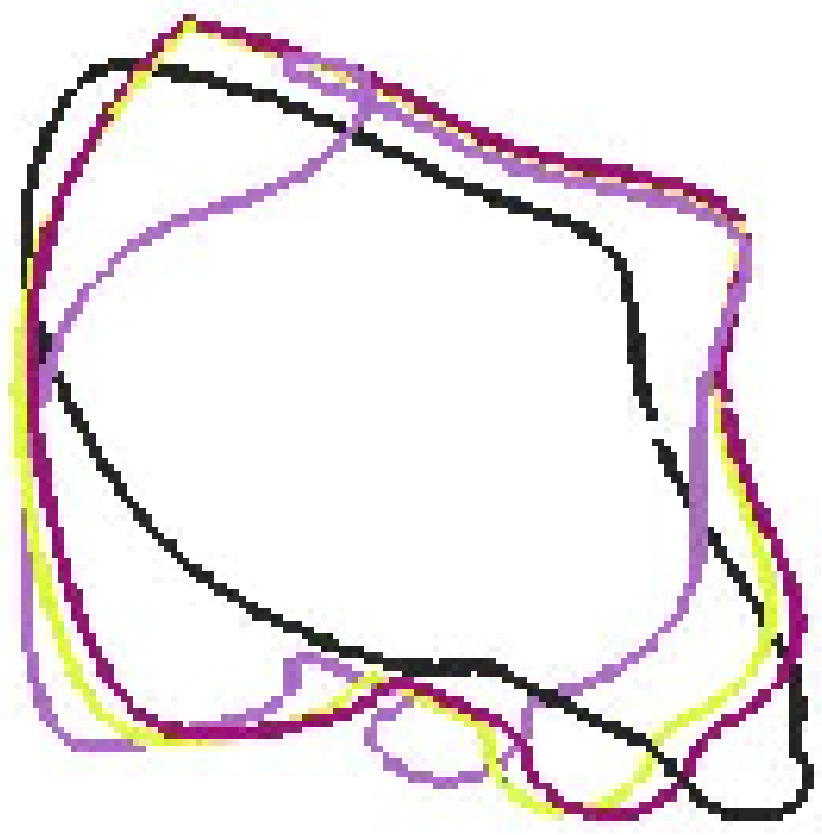}
\caption{{\bf Conservation of entanglement}. The figure shows the
entanglement due to helicity conservation. Each curve results from
the intersection of u and v level surfaces. It
displays few curves at times equal 0.3, 0.5 and 0.8 from left to right. 
The values of
the level surfaces are different but at any instant of time there
is the same number of linked curves.} \label{fig2}
\end{figure}

In Figure \ref{fig1} we plot the evolution of two level surfaces
for a domain size $L = 5$ and we have used 100 grid points in each
direction. They correspond to the values $u = -0.7$ and $v =
-0.1$. Those surfaces like them have a non-trivial intersection.
One can use a marching cubes algorithm \cite{r21} to get the
intersections. In Figure \ref{fig2} we have plotted the
intersection of few pairs of $(u, v)$ level surfaces at different
times. We can see that, as we have shown theoretically, they are
linked. It is possible to find linked curves in all the instant of
times as far we have enough numerical precision. The system
eventually will decay to the equilibrium values fixed by the
boundary conditions, but it will do that keeping the linking
number constant. In conclusion, we have presented a new global
approach to investigate how organizing centres in excitable media
persist or disappear in three dimensions. In particular, we have
defined the helicity of an excitable medium as a quantity that
describes the global linkage of the intersections of the level
surfaces of the scalar fields in the medium. We have studied how
the helicity is conserved or lost through the walls of the medium
and shown that these behaviours are dominated by the boundary
conditions, so the distortion of these conditions could lead to
the disappearance of the structures.

We thank S. Betel\'u for discussions on parts of this research. 
The authors thank support from the Spanish Ministerio de
Educaci\'on y Ciencia under project ESP2007-66542-C04-03.

\newpage

\end{document}